\documentclass[%
preprintnumbers,
 amsmath,amssymb,
 aps, physrev, 
 twocolumn
]{revtex4-2}

\usepackage{gensymb}
\usepackage{subfigure}
\usepackage{graphicx}% Include figure files
\usepackage{dcolumn}% Align table columns on decimal point
\usepackage{bm}% bold math
\usepackage{xcolor}
\usepackage{tikz}
\usepackage{bbm}
\usetikzlibrary{decorations.pathmorphing}
\usetikzlibrary{decorations.markings}% bold math

\begin{document}

% observed limits
% observed:

% ATLAS: $m_W$ via $p_T^\ell$
% 95% CL interval: [-0.118, 0.036]
% 68% CL interval: [-0.080, -0.002]
% central value: -0.041

% CMS: $m_W$ via $p_T^\ell$
% 95% CL interval: [-0.002, 0.162]
% 68% CL interval: [0.038, 0.122]
% central value: 0.080

% LHC combination
% 95% CL interval: [-0.041, 0.071]
% 68% CL interval: [-0.013, 0.044]
% central value: 0.015

% CDF: $m_W$ via $m_T^W$
% 95% CL interval: [1.079, 1.979]
% 68% CL interval: [1.291, 1.756]
% central value: 1.521

\preprint{IPPP/25/43}

\title{\textbf{Sensitivity of $W$-boson measurements to low-mass right-handed neutrinos} }% 

\author{Rodrigo Alonso}
\author{Michael Spannowsky}
\affiliation
{Institute of Particle Physics Phenomenology, University of Durham, Durham, UK}

 \author{Sam Bates}
\author{Chris Hays}
 \email{Contact author: chris.hays@physics.ox.ac.uk}
\affiliation{Department of Physics, University of Oxford, Oxford, UK}

\author{Chris Pollard}
\affiliation{Department of Physics, University of Warwick, Coventry, UK}

\date{\today}

\begin{abstract}
A low-mass right-handed neutrino could interact with electroweak bosons via mixing, a mediator particle, or loop corrections.  Using an effective field theory, we determine constraints on these interactions from $W$-boson measurements at hadron colliders.  Due to the difference in the initial states at the Tevatron and the LHC, $W$-boson decays to a right-handed neutrino would artificially 
increase the mass measured at the Tevatron while only affecting the difference 
between $W^+$ and $W^-$ mass measurements at the LHC.  Measurements from CDF and the 
LHC are used to infer the corresponding parameter values, which are 
found to be inconsistent between the two.  The LHC experiments can improve 
sensitivity to these interactions by measuring the cosine of the helicity angle 
using $W$ bosons produced with transverse momentum above $\approx 50$~GeV.  
\end{abstract}

\maketitle

%\tableofcontents

\section{\label{sec:intro}Introduction}
The discovery of the Higgs boson~\cite{HIGG-2012-27,CMS-CMS-00-002} 
completed the observed particle content of the Standard Model (SM), 
and subsequent measurements~\cite{CMS-HIG-22-001,HIGG-2023-11} are consistent with its prediction that all electrically charged particles and the $Z$ boson receive their mass from the vacuum expectation value ($v$) in the Higgs mechanism~\cite{PhysRevLett.13.321,PhysRevLett.13.508,PhysRevLett.13.585}. 
However, the source of neutrino mass remains elusive, and there is no evidence that it is due to a Dirac or Majorana term, or both. Introducing a singlet right-handed neutrino $\nu_R$ would allow for a Dirac mass term, in which case the new field would provide a new spin state rather than a new particle. The SM neutrinos could instead attain a Majorana mass via the sole dimension-5 operator~\cite{PhysRevLett.44.912} in the SM effective field theory (SMEFT)~\footnote{See~\cite{Brivio_2019,Isidori_2024} for reviews of the SMEFT}.  This operator could be generated in the renormalizable Seesaw mechanism~\cite{MINKOWSKI1977,PhysRevLett.44.912,yanagida,Gell-Mann:1979vob} by the exchange of a heavy neutrino ($m_N>v$) or a heavy scalar triplet.  The presence of a singlet $\nu_R$ field can thus accommodate both Majorana and Dirac masses for active neutrinos.

The introduction of $\nu_R$ fields need not be tied to active neutrino masses. Thus, in this work, we consider the existence of $\nu_R$ states leading to one or more low-mass neutrinos ($m_N\lesssim 1$~GeV).  Such neutrinos could be the source of the observed dark matter in the universe~\cite{BOYARSKY2019} and could give a rich phenomenology at particle colliders~\cite{Alcaide:2019pnf,Butterworth_2019,Beltran:2025ilg,Mitra:2022nri,Mitra:2024ebr}, particularly if they are associated with other new states at the electroweak scale.  We focus on possible new effective interactions of the $\nu_R$ states with $W$ bosons, which could be generated by a small mixing of this state with active neutrinos, the exchange of another mediator particle, or loop corrections.

Regardless of its origin, this interaction will produce the phenomenologically relevant decay $W \rightarrow \ell \nu_R$, where $\ell$ is an electron or muon. This decay would affect measurements of the $W$ boson mass at the Tevatron and LHC, and could be further probed through measurements of angular distributions in the $W$ boson decay.

We determine constraints from existing and potential measurements using a $\nu_R$-augmented SM effective field theory, the $\nu$SMEFT~\cite{DELAGUILA2009}, which captures all the aforementioned 
possibilities for the generation of the interaction. The relevant Lagrangian terms with dimension-6 operators are  
\begin{eqnarray}
    \Delta {\cal{L}} & =  \frac{c_{L N H}^i}{\Lambda^2} \bar{L}_i \nu_R \tilde{H} H^{\dag}H + \frac{c_{H N e}^i}{\Lambda^2} \bar{\nu}_R\gamma^\mu e_{iR}\tilde H^{\dag} iD_\mu H  \nonumber \\ 
    & + \frac{c_{NW}^i}{\Lambda^2} \bar{L}_i \sigma^{\mu\nu} \nu_R \sigma_I \tilde{H} W_{\mu\nu}^I + {\textrm{h.c.}}, 
\end{eqnarray}
\noindent
where $c_{LNH}^i$, $c_{HNe}^i$, and $c_{NW}^i$ are effective couplings, $i = e,\mu,\tau$, the mass scale of the new particles is $\Lambda$, $L$ is the left-handed lepton doublet, $H$ is the Higgs doublet, and $\tilde{H} = i\sigma_2 H^*$.  
Assuming real couplings, a non-zero $c_{HNe}^i$ causes a parity-violating V+A interaction, whereas the other two couplings cause chirality-flipping interactions~\footnote{A renormalizable Yukawa term $Y^i\bar{L}_i \nu_R \tilde{H}$ (as well as the operator with coefficient $c_{LNH}^i$) generates neutrino mixing and produces the opposite-chirality interactions that we do not consider in our phenomenological study. This mixing is constrained to be small ($Yv/m_N\ll 1$), see e.g.~\cite{PhysRevD.100.073011,Fernandez-Martinez:2023phj}. The $c_{NW}^i$ term is loop-generated in weakly coupled renormalizable UV completions, and is thus small in general.}.

Focusing on the V+A interaction, a Dirac neutrino with a mass larger than 1 MeV will decay into an electron, positron, and a neutrino with a lifetime
\begin{equation}
     \Gamma = \frac{m_N^5G_F^2}{192\pi^3}\frac{ v^4 \sum_i |c_{HNe}^i|^2}{4\Lambda^4}.
\end{equation}
while a Majorana neutrino has twice this decay rate due to the CP-conjugate channel.

Perturbative unitarity provides an upper bound on the coupling of order
\begin{align}
    \frac{c_{HNe} E^2}{\Lambda^2}&\lesssim 4\pi,
\end{align}
which also gives an upper bound on the mass scale at which new states should appear. For consistency, this scale should be greater than $v$, so the perturbative bound is here
taken to be 
\begin{align}
    \left|\frac{c_{HNe}}{\Lambda^2}\right|\leq\frac{4\pi}{v^2}.
\end{align} 

For a value of $c_{HNe}^i/\Lambda^2=1/v^2$ and a mass of 100 
MeV, the vast majority of neutrinos decay outside the fiducial 
volume of a typical detector ($\gamma c\tau\approx 100$~m).  
These decays can thus be collected with a standard $W$ boson 
selection at a collider.  We find that existing $W$ boson 
measurements have sensitivity to values of $c_{HNe}^i$ as low 
as $0.1/v^2$ and are valid for neutrino masses up to $\approx 
1$~GeV.  Additional $\nu$SMEFT operators could affect these 
results by reducing the neutrino lifetime, in particular by 
allowing $\nu_R \to \nu \gamma$ decay~\cite{PhysRevD.92.093002}.  
Here, we assume the associated Wilson coefficients are small enough for $\nu_R$ to dominantly decay outside the detector.

Prior measurements and searches constrain $W$ boson couplings to low-mass right-handed neutrinos. The electron angular distributions in nuclear beta decay constrain couplings to electron neutrinos~\cite{Falkowski:2020pma}, and those in muon decay additionally constrain couplings to muon neutrinos~\cite{10.21468/SciPostPhysProc.5.006}.  The $\tau$-lepton lifetime constrains couplings to all neutrino flavours, and direct searches for GeV-scale neutrinos at ATLAS~\cite{ATLAS:2025uah} and CMS~\cite{CMS:2022fut,CMS:2023jqi,CMS:2024hik,2025570} constrain couplings to electron and muon neutrinos.  All constraints are at the few-percent level 
relative to the coupling to left-handed neutrinos.  Upper bounds have also been placed on $\nu_L$--$\nu_R$ mixing~\cite{PhysRevD.100.073011}.  Our results rely 
on a potential V+A interaction in the decay of $W$ bosons and are given in terms of the coupling $c_{HNe}^i$.  We consider scenarios where there is one non-zero coefficient $c_{HNe}$ ($i=e$) or $c_{HN\mu}$ ($i=\mu$), or where the electron and muon coefficients are the same value, $c_{HN\ell}$.

This paper is structured as follows: 
Sec.~\ref{sec:wmeasurements} quantifies the impact of $W \rightarrow \ell \nu_R$ decays on existing and potential Tevatron and LHC $W$ boson measurements in terms of $c_{HNe}^i$; Sec.~\ref{sec:other} discusses $\nu_L$--$\nu_R$ mixing and presents the competitive constraints on $c_{HNe}^i$ from pion and muon decay; and Sec.~\ref{sec:conclusions} summarizes our conclusions.

\section{$W$ boson measurements}
\label{sec:wmeasurements}
Measurements of the $W$ boson at the Tevatron and 
LHC~\cite{LHCTeV:2023zkn,CMS:2024lrd,ATLAS:2024erm,Aaboud:2017svj,LHCb:2021bjt,CDF:2022hxs,D0:2013jba,Abazov:2009cp} are performed assuming that its 
production and decay follow the predictions of the
Standard Model. Decays to right-handed neutrinos
would affect these measurements, with a different
impact at each collider due to differences between
up-quark, down-quark, and sea-quark momentum
distributions in the proton.  We quantify the
$c_{HNe}^i$ dependence of existing and potential
measurements of the $W$ boson mass ($m_W$), 
cross-section, and cosine of the helicity angle in 
the $W$ boson decay~\footnote{The $W$ boson width 
is also affected by $c_{HNe}^i$, however its current 
value of $2014 \pm 50$~MeV~\cite{ParticleDataGroup:2024cfk} corresponds to a 
95\% confidence-level constraint of $c_{HNe}^i < 
1.6$ for $\Lambda = v$~\cite{Butterworth_2019}.  
This is significantly weaker than the constraints 
we consider.  Interpreting the recent ATLAS width 
measurement~\cite{ATLAS:2024erm} in terms of $c_{HNe}^i$ would 
give $c_{HNe}^i = 1.4 \pm 0.6$, in agreement with 
the value we obtain from the CDF $m_W$ measurement but in tension with that obtained from LHC measurements of $m_{W^+} - m_{W^-}$.}.

\subsection{CDF $m_W$ measurement}

The CDF measurement of the $W$ boson mass~\cite{CDF:2022hxs} 
was performed using template fits to distributions of final-state kinematic quantities projected transverse to the 
beamline.  The templates were produced at different values of 
$m_W$, assuming SM production and decay. The measured mass is 
76 MeV higher than the SM prediction, a difference that cannot 
be explained 
currently~\cite{Biswas:2024wbz,Agashe:2024owh,Alonso:2024pmq,deGiorgi:2023wjh,Giarnetti:2023dcr,Chowdhury:2023uyd,Ahriche:2023hho,Agashe:2023itp}.  

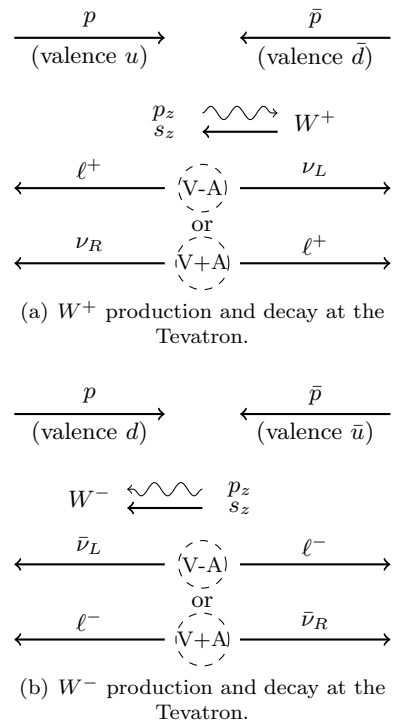
\begin{figure}[ht!]
\centering
\subfigure[$W^+$ production and decay at the Tevatron. ]{
\begin{tikzpicture}
    \draw [thick,->] (-0.5,4.75) --(1.5,4.75);
    \draw [thick,<-] (2.5,4.75) --(4.5,4.75);
    \draw [thick] (0.5,5) node {$p$};
    \draw [thick] (0.5,4.5) node {(valence $u$)};
    \draw [thick] (3.5,5) node {$\bar{p}$};
    \draw [thick] (3.5,4.5) node {(valence $\bar{d}$)};
 \draw [decorate,decoration=snake,->]  (2.,3.75) -- (3.,3.75) node [anchor=east]{};
     \draw [thick,<-] (2.,3.5) --(3.,3.5);

 \draw [thick] (3.5,3.65) node {$W^+$};
 \draw [thick] (1.5,3.75) node {$p_z$};
 \draw [thick] (1.5,3.5) node {$s_z$};
    \draw [thick,<-] (-0.5,2.75) --(1.5,2.75);
    \draw [thick,->] (2.5,2.75) --(4.5,2.75);
    \draw [thick] (2,2.75) node  {\footnotesize V-A}; %R.A. my addition, remove if desired
    \draw [dashed] (2,2.75) circle (9pt); %R.A. my addition, remove if desired
    \draw [thick] (3.5,3) node {$\nu_L$};
    \draw [thick] (0.5,3) node {$\ell^+$};
    \draw [thick] (2.,2.25) node {or};
    \draw [thick,<-] (-0.5,1.75) --(1.5,1.75);
    \draw [thick,->] (2.5,1.75) --(4.5,1.75);
    \draw [thick] (2,1.75) node  {\footnotesize V+A}; %R.A. my addition, remove if desired
    \draw [dashed] (2,1.75) circle (10pt);%R.A. my addition, remove if desired
    \draw [thick] (0.5,2) node {$\nu_R$};
    \draw [thick] (3.5,2) node {$\ell^+$};
   \end{tikzpicture}
   }
   \vskip 0.1in
   \subfigure[$W^-$ production and decay at the Tevatron. ]{
\begin{tikzpicture}
    \draw [thick,->] (-0.5,4.75) --(1.5,4.75);
    \draw [thick,<-] (2.5,4.75) --(4.5,4.75);
    \draw [thick] (0.5,5) node {$p$};
    \draw [thick] (0.5,4.5) node {(valence $d$)};
    \draw [thick] (3.5,5) node {$\bar{p}$};
    \draw [thick] (3.5,4.5) node {(valence $\bar{u}$)};
 \draw [decorate,decoration=snake,->]  (2.,3.75) -- (1.,3.75) node [anchor=east]{};
    \draw [thick,->] (2.,3.5) --(1.,3.5);
 \draw [thick] (0.5,3.65) node {$W^-$};
 \draw [thick] (2.5,3.75) node {$p_z$};
 \draw [thick] (2.5,3.5) node {$s_z$};
    \draw [thick,<-] (-0.5,2.75) --(1.5,2.75);
    \draw [thick,->] (2.5,2.75) --(4.5,2.75);
    \draw [thick] (2,2.75) node  {\footnotesize V-A}; %R.A. my addition, remove if desired
    \draw [dashed] (2,2.75) circle (9pt); %R.A. my addition, remove if desired
    \draw [thick] (0.5,3) node {$\bar{\nu}_L$};
    \draw [thick] (3.5,3) node {$\ell^-$};
    \draw [thick] (2.,2.25) node {or};
    \draw [thick,<-] (-0.5,1.75) --(1.5,1.75);
    \draw [thick,->] (2.5,1.75) --(4.5,1.75);
    \draw [thick] (3.5,2) node {$\bar\nu_R$};
    \draw [thick] (0.5,2) node {$\ell^-$};
    \draw [thick] (2,1.75) node  {\footnotesize V+A}; %R.A. my addition, remove if desired
    \draw [dashed] (2,1.75) circle (10pt);%R.A. my addition, remove if desired

   \end{tikzpicture}
   }
   \caption{The production and decay of $W^{\pm}$ bosons at the Tevatron for 
   the final state containing either $\nu_L$ or $\nu_R$.  On average the $W^\pm$ boson $p_z$ is in the direction of the interacting \mbox{(anti-)}$u$ quark, since $\langle p_z^u \rangle > \langle p_z^d \rangle$. Due to the V$-$A interaction, the initial spin is along the anti-proton direction, so the left-handed neutrino is produced in the direction of the $W$ boson. In the case of $\nu_R$, the $W$-boson interaction with the charged lepton is V+A, so the direction is reversed.}
   \label{fig:tevmw}
\end{figure}

$W$ boson decays to right-handed neutrinos would artificially increase the measured value of $m_W$ at the Tevatron due to the effects of parton momentum distributions, lepton decay angles, and detector acceptance.  On average, the valence up quark in the proton has a higher momentum than the valence down quark.  This leads to an average momentum along the proton direction for $W^+$ bosons, and vice versa for $W^-$ bosons (see Fig.~\ref{fig:tevmw}).  For V$-$A (V+A) decays, the direction of the charged lepton is along (opposite to) that of the $W^+$ boson.  Consequently, the charged lepton has a lower average $|p_z|$ for the SM decay than for the decay to $\nu_R$.  

\begin{figure}[ht!]
\subfigure[]{
\includegraphics[width=0.49\textwidth]{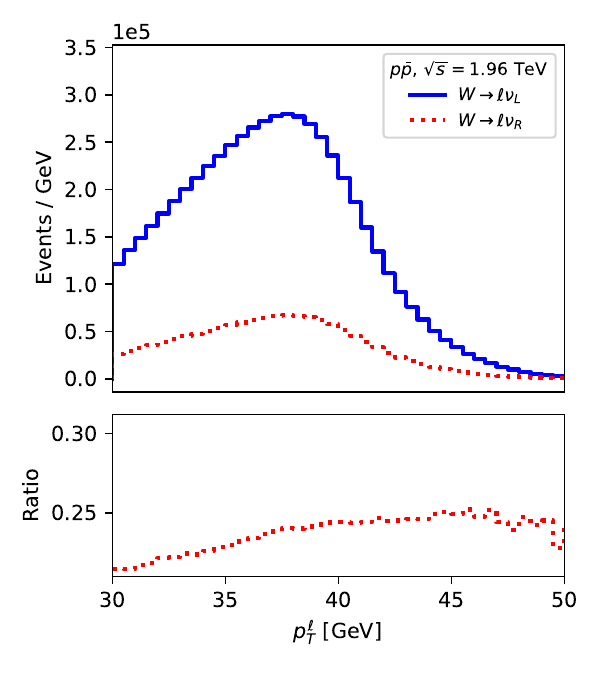}%
}
\subfigure[]{
\includegraphics[width=0.49\textwidth]{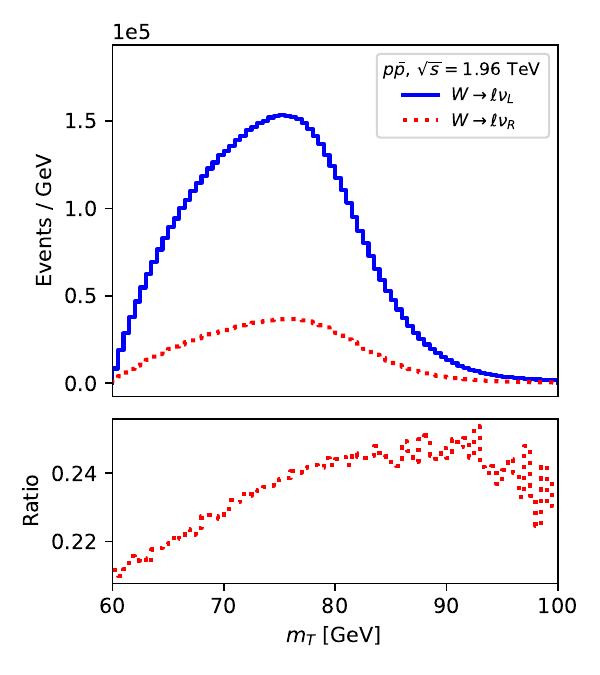}%
}
\caption{The lepton $p_T$ (a) and $m_T$ (b) distributions for SM production and decay of the $W$ boson ($W\to \ell\nu_L$) at CDF, and for the $W$ boson decaying to a right-handed neutrino ($W\to \ell\nu_R$) with effective coupling $c_{HNe}/\Lambda^2 = 1/v^2$. }
\label{fig:cdfptmt}
\end{figure}

The lepton $p_z$ distribution affects the measured $m_W$ through detector acceptance. The CDF measurement used charged leptons with absolute pseudorapidity ($|\eta|$) less than 1.  Increasing $p_z$ increases the fraction of accepted leptons that decay perpendicular to the beamline. These decays 
correspond to the peak transverse momentum. The resulting $p_T$ and $m_T$ distributions, shown in Fig.~\ref{fig:cdfptmt}, lead to a larger measured $m_W$ if one assumes SM decay.

To determine the bias in $m_W$ as a function of $c_{HNe}/\Lambda^2$, we separately generate leading-order $W\to e\nu_L$ SM decays and $W\to e\nu_R$ decays with $c_{HNe}=1$ and $\Lambda=v$, for $\sqrt{s}=1.96$ TeV $p\bar{p}$ collisions using the {\textsc{MadGraph5\_aMC@NLO}} 
generator~\cite{Alwall:2014hca}.  We apply the CDF event selection and model the detector effects by approximately reproducing the measured kinematic distributions. The inclusion of higher orders and an improved detector modelling would affect the details of these distributions, but would 
not significantly affect the difference between SM decays and decays to $\ell\nu_R$.

We estimate the measured $m_W$ bias from a non-zero $c_{HNe}$ using a template fit to the $m_T$ distribution (Fig.~\ref{fig:cdfptmt}), where the simulated data contain the sum of SM decays and decays to $\nu_R$, and the template 
contains SM decays only. The total yields of the simulated data and the templates are normalised to the number of events observed by CDF. The $m_W$ bias is determined as a function of the signal strength $\mu$, defined as $N_{sig} = \mu N_{gen}$, where $N_{sig}$ is the number of signal events in the simulated data and $N_{gen}$ is the number of events for the generated parameter values $c_{HNe}=1$ and $\Lambda=v$.  
The measured $m_W$ value corresponds to $|c_{HN\ell}| v^2/\Lambda^2 = 1.2 \pm 0.1$, near the perturbativity limit for a new-physics scale equal to the vacuum expectation value.

A caveat to our estimated bias on $m_W$ is the impact a non-zero $c_{HN\ell}$ value would have on the charge asymmetry of the $W$ boson cross-section as a function of $W$ boson rapidity. Tevatron charge asymmetry measurements~\cite{D0:2014kma,D0:2013xqc,CDF:2009cjw} 
are used in fits for parton distribution functions, and these fits would be affected by a non-zero $c_{HN\ell}$. To estimate the impact on the $m_W$ bias, we reweight the $u/d$ ratio in the generated events such that the charge asymmetry is unchanged for the SM distribution. The reweighting suppresses the $m_W$ bias by about 30\%, so one would have to divide the estimated $\mu$ value by about 0.7 to account for this effect. In practice, the charge asymmetry measurements are just one component of a global PDF fit, and the obtained $\mu$ value is inconsistent with those derived from ATLAS and CMS below, so we do not attempt to refine our $\mu$ estimate.

\subsection{ATLAS and CMS $m_W$ measurements}

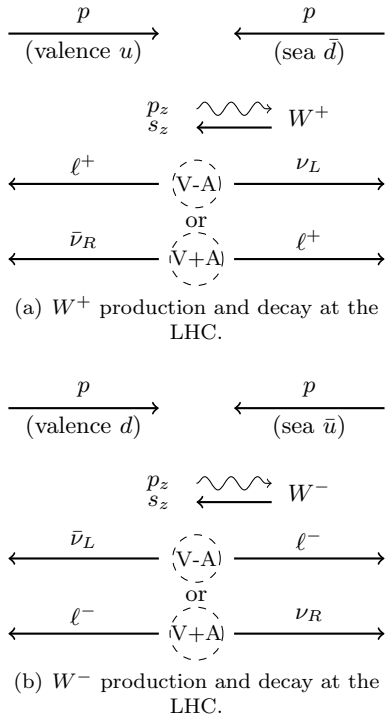
\begin{figure}[ht!]
\centering
\subfigure[$W^+$ production and decay at the LHC. ]{
\begin{tikzpicture}
    \draw [thick,->] (-0.5,4.75) --(1.5,4.75);
    \draw [thick,<-] (2.5,4.75) --(4.5,4.75);
    \draw [thick] (0.5,5) node {$p$};
    \draw [thick] (0.5,4.5) node {(valence $u$)};
    \draw [thick] (3.5,5) node {$p$};
    \draw [thick] (3.5,4.5) node {(sea $\bar{d}$)};
% W 
 \draw [decorate,decoration=snake,->]  (2.,3.75) -- (3.,3.75) node [anchor=east]{};
     \draw [thick,<-] (2.,3.5) --(3.,3.5);

 \draw [thick] (3.5,3.65) node {$W^+$};
 \draw [thick] (1.5,3.75) node {$p_z$};
 \draw [thick] (1.5,3.5) node {$s_z$};
% vL
    \draw [thick,<-] (-0.5,2.75) --(1.5,2.75);
    \draw [thick,->] (2.5,2.75) --(4.5,2.75);
    \draw [thick] (3.5,3) node {$\nu_L$};
    \draw [thick] (0.5,3) node {$\ell^+$};
    \draw [thick] (2.,2.25) node {or};
    \draw [thick] (2,2.75) node  {\footnotesize V-A}; %R.A. my addition, remove if desired
    \draw [dashed] (2,2.75) circle (9pt); %R.A. my addition, remove if desired
% vR
    \draw [thick,<-] (-0.5,1.75) --(1.5,1.75);
    \draw [thick,->] (2.5,1.75) --(4.5,1.75);
    \draw [thick] (0.5,2) node {$\bar\nu_R$};
    \draw [thick] (3.5,2) node {$\ell^+$};
    \draw [thick] (2,1.75) node  {\footnotesize V+A}; %R.A. my addition, remove if desired
    \draw [dashed] (2,1.75) circle (10pt);%R.A. my addition, remove if desired
   \end{tikzpicture}
}
\vskip 0.1in
\subfigure[$W^-$ production and decay at the LHC.]{
\begin{tikzpicture}
    \draw [thick,->] (-0.5,4.75) --(1.5,4.75);
    \draw [thick,<-] (2.5,4.75) --(4.5,4.75);
    \draw [thick] (0.5,5) node {$p$};
    \draw [thick] (0.5,4.5) node {(valence $d$)};
    \draw [thick] (3.5,5) node {$p$};
    \draw [thick] (3.5,4.5) node {(sea $\bar{u}$)};
% W 
 \draw [decorate,decoration=snake,->]  (2.,3.75) -- (3.,3.75) node [anchor=east]{};
     \draw [thick,<-] (2.,3.5) --(3.,3.5);

 \draw [thick] (3.5,3.65) node {$W^-$};
 \draw [thick] (1.5,3.75) node {$p_z$};
 \draw [thick] (1.5,3.5) node {$s_z$};
% vL
    \draw [thick,<-] (-0.5,2.75) --(1.5,2.75);
    \draw [thick,->] (2.5,2.75) --(4.5,2.75);
    \draw [thick] (3.5,3) node {$\ell^-$};
    \draw [thick] (0.5,3) node {$\bar{\nu}_L$};
    \draw [thick] (2,2.75) node  {\footnotesize V-A}; %R.A. my addition, remove if desired
    \draw [dashed] (2,2.75) circle (9pt); %R.A. my addition, remove if desired
    \draw [thick] (2.,2.25) node {or};
% vR
    \draw [thick,<-] (-0.5,1.75) --(1.5,1.75);
    \draw [thick,->] (2.5,1.75) --(4.5,1.75);
    \draw [thick] (0.5,2) node {$\ell^-$};
    \draw [thick] (3.5,2) node {$\nu_R$};
    \draw [thick] (2,1.75) node  {\footnotesize V+A}; %R.A. my addition, remove if desired
    \draw [dashed] (2,1.75) circle (10pt);%R.A. my addition, remove if desired
   \end{tikzpicture}
}
 
   \caption{The production and decay of $W^\pm$ bosons at the LHC for the final state containing either $\nu_L$ or $\nu_R$.  The $W^\pm$ boson momentum projected along the $z$-axis is in the direction of the proton contributing a valence quark.  The decay to $\nu_L$ is V-A, so the $\nu_L$ direction is opposite to that of the $W^+$ boson.  In the case of $\nu_R$, the decay is V+A, and the direction is reversed.}
   \label{fig:lhcmw}
\end{figure}

\begin{figure}[ht!]
\subfigure[]{
\includegraphics[width=0.47\textwidth]{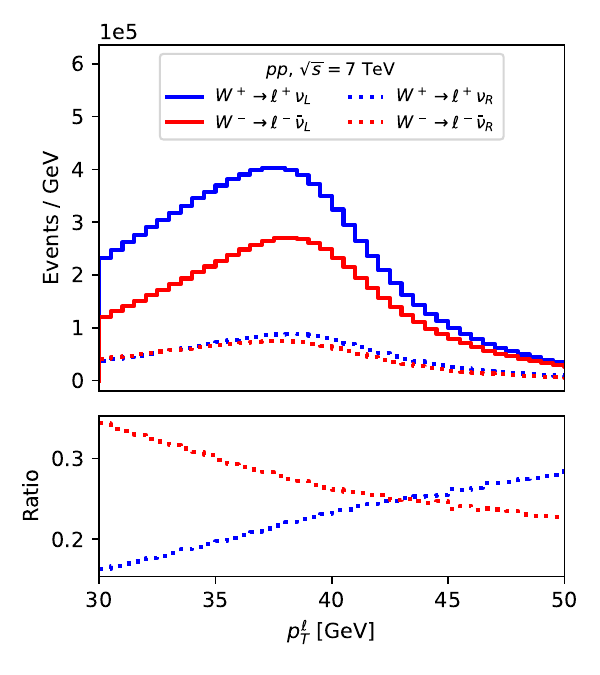}%
}
\subfigure[]{
\includegraphics[width=0.47\textwidth]{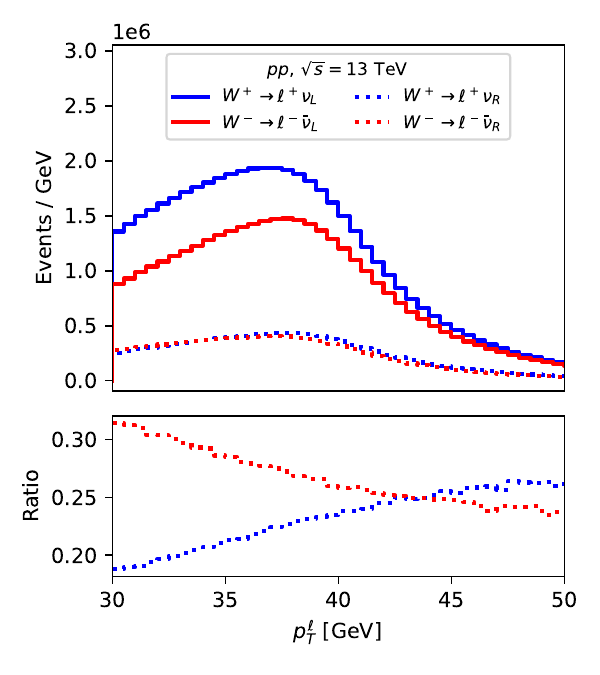}%
}
\caption{The lepton $p_T$ distribution for SM production and decay of the $W^+$ and $W^-$ bosons to a given lepton ($W^\pm \to \ell\nu_L$) at the LHC with a center-of-mass energy of (a) 7 TeV or (b) 13 TeV, and for the corresponding decays to a right-handed neutrino ($W^\pm \to \ell\nu_R$) with effective coupling $c_{HNe}/\Lambda^2 = 1/v^2$.}
\label{fig:lhcpt}
\end{figure}

The $m_W$ measurements with the ATLAS~\cite{ATLAS:2024erm,Aaboud:2017svj} and CMS~\cite{CMS:2024lrd} experiments are sensitive to the $W\to \ell\nu_R$ decay dominantly through the difference between the $W^+$ and $W^-$ boson mass measurements. $W$ bosons tend to be produced with $p_z$ in the direction of the interacting valence quark, so $W^-$ bosons are generally produced at the 
LHC in the direction of the $d$ valence quark, rather than the $\bar{u}$ quark at the Tevatron.  This is represented pictorially in Fig. \ref{fig:lhcmw}, and the impact on the $p_T^{\ell}$ distribution is shown in Fig. \ref{fig:lhcpt}.

\begin{figure}[ht!]
\subfigure{
\includegraphics[width=0.49\textwidth]{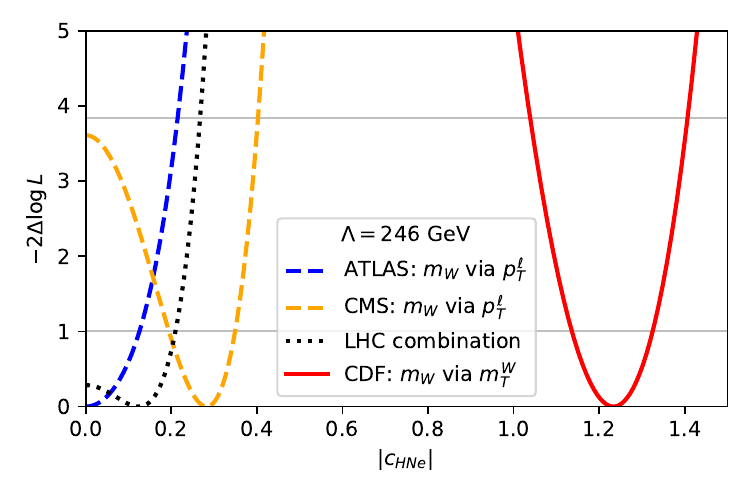}%
}
\caption{The negative log likelihood as a function of $|c_{HNe}|$ inferred from the CDF $m_W$ measurement and the ATLAS and CMS $m_{W^+} - m_{W^-}$ measurements. Also shown is the likelihood from a combination of the LHC measurements, neglecting correlations. }
\label{fig:llh}
\end{figure}

The measurement biases at ATLAS and CMS are evaluated in the same way as for CDF, i.e. by applying momentum resolution functions to generated $W\to \ell\nu_L$ and $W\to \ell\nu_R$ decays at the appropriate center-of-mass energy, and performing a fit to the $p_T^{\ell}$ distribution for the $W^+$ and $W^-$ boson masses as a function of $c_{HN\ell}/\Lambda^2$. The results are compared to the measured mass differences, $m_{W^+} - m_{W^-} = -29.2 \pm 28.0$ and $57.0 \pm 30.3$ MeV from ATLAS and CMS, respectively. The corresponding best-fit values for $|c_{HN\ell}|$ are $0^{+0.13}_{-0}$ and $0.28 \pm 0.09$ for 
$\Lambda=v$.  The mass differences at ATLAS and CMS have only a small correlation through the subleading PDF uncertainty. Neglecting this correlation gives a combined value of $|c_{HN\ell}| v^2/\Lambda^2 = 0.12^{+0.15}_{-0.12}$, 
or a 95\% confidence-level interval of $[0, 0.27]$.  For $\Lambda=1$ TeV the value is $|c_{HN\ell}| = 2^{+2.5}_{-2}$.  These values are incompatible with those obtained from the CDF $m_W$ result, as shown in Fig.\ref{fig:llh}.  

The CMS $W$ boson mass measurement constrains PDF uncertainties in situ using the total event yields of $W^+ \to \mu^+ \nu_L$ and $W^- \to \mu^- \bar{\nu}_L$. These event yields would also be affected by the presence of $W \to \mu^- \nu_R$ decays due to the different acceptances of these decays. To understand the sensitivity of the event yield to 
$c_{HN\ell}/\Lambda^2$, we use the ATLAS $W$ boson cross section measurements at 7 TeV to fit for $c_{HN\ell}/\Lambda^2$, and find an uncertainty of 
$\approx 0.2$ on $c_{HN\ell}$ for $\Lambda = v$.  This is approximately twice the uncertainty obtained from $m_{W^+} - m_{W^-}$, so we expect the in situ PDF constraint to not significantly affect the extracted constraint on $c_{HN\ell}^2 v^4/\Lambda^4$.

\subsection{$W$ boson angular measurement}
\label{sec:costheta}

The most directly sensitive observable to V+A decay is the lepton decay angle in the $W$ boson rest frame, $\theta^*$.  Because of the unobserved neutrino, this angle can only be determined up to a two-fold ambiguity, and the resulting distribution is symmetric for V$-$A and V+A decay. However, the ambiguity can be partially resolved if one considers $W$ bosons boosted in the transverse plane~\cite{CMS:2011kaj,ATLAS:2012au}. At high $p_T^W$, the projection of the lepton onto the $W$ boson in the transverse plane~\cite{CMS:2011kaj}, 
\begin{equation}
L_p \equiv \frac{\vec{p}_T^{~\ell} \cdot \vec{p}_T^{~W}}{|p_T^W|^2},
\end{equation}
\noindent
allows $\theta^*$ to be approximated by $\cos\theta^* \approx 2 L_p -1$. Projection variables have been used to measure the helicity fractions of the $W$-boson, and can alternatively be used to constrain V+A interactions in its decay directly.  

\begin{figure}[ht!]
\includegraphics[width=0.49\textwidth]{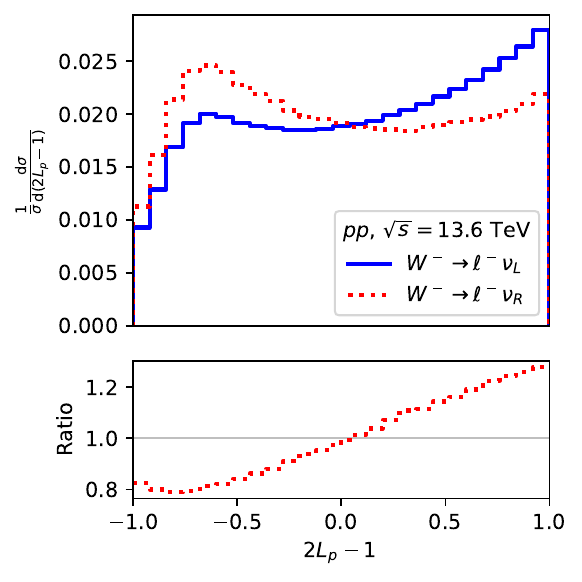}%
\caption{The $2L_p - 1$ distribution for SM production and 
decay of the $W^-$ boson to a given lepton ($W^- \to \ell^-\nu_L$) at the LHC with a center-of-mass energy of 13.6 TeV, and for the corresponding decay to a right-handed neutrino ($W^- \to \ell^-\nu_R$). The $W^+$ distributions are the same when plotted as a function of $-(2L_p - 1)$.}
\label{fig:lhcang}
\end{figure}

Figure~\ref{fig:lhcang} shows the $2 L_p - 1$ distribution for $W^-$ bosons decaying to left-handed or right-handed neutrinos for events with $p_T^W >50$~GeV, lepton $p_T>25$~GeV, and $m_T > 50$~GeV.  Given the good discrimination between the two cases, we estimate the expected $\nu_R$ sensitivity of a $2L_p - 1$ measurement of $W^\pm$ decays in the Run 3 data (300 fb$^{-1}$ of integrated luminosity at $\sqrt{s} = 13.6$~TeV).  In this dataset, the statistical uncertainty will be below the percent level in the discriminating region of $2L_p - 1$.  We thus expect systematic uncertainties to be relevant and do not project to the larger HL-LHC dataset.  

Assuming equal contributions of statistical and systematic uncertainties, the expected 95\% C.L. constraint from Run 3 data is $|c_{HNe}| < 0.08$ (and the same for $|c_{HN\mu}|$).  These constraints are tighter than those from existing measurements of $m_{W^+} - m_{W^-}$, though future measurements of the mass difference should give competitive sensitivity.

\section{Other constraints}
\label{sec:other}
A non-zero $c_{HN\ell}$ would impact various other phenomena, of which we explicitly consider pion decay, nuclear beta decay, and muon decay~\footnote{We only discuss the most constraining measurements in the mass range 0.1--100 MeV.  We have also investigated the constraint arising from $e^+e^- \to W^+W^-$ production at LEP~\cite{ALEPH:2013dgf}, and find it to be weaker than the others ($|c_{HN\ell}| < 0.51$).}.  To compare limits arising from these measurements, we determine the effective mixing caused by $c_{HN\ell}$ between the active neutrinos and a fourth singlet neutrino. The leptonic $W$-boson couplings are
\begin{align}
    \mathcal L_{CC}^\ell =- \frac{g}{\sqrt 2} W_\mu^-\left(\bar \nu_L^i \gamma_\mu e_L^i+\frac{c_{HNe}^i v^2}{2\Lambda^2}\bar \nu_R \gamma_\mu e_R^i\right)+h.c.,
\end{align}
which when taken as the couplings of an array of four neutrinos with three charged leptons gives a mixing of
\begin{align}
    U_{i 4}=\frac{c_{HNe}^iv^2}{2\Lambda^2}.
\end{align}
Using this relation, Ref.~\cite{Fernandez-Martinez:2023phj} translated various 
mixing bounds to bounds on $c^i_{HNe}$. From their analysis, we conclude that various other experimental constraints are stronger than the bounds from $W$-boson measurements for neutrino masses higher than a few MeV. 

For the $c_{HNe}$ coupling, the strongest bound 
transitions from $\pi\to \nu +e$ 
searches~\cite{PhysRevLett.115.071801} interpreted as 
constraints on $U_{e4}$~\cite{PhysRevD.100.073011} to 
nuclear $\beta$ decays~\cite{Falkowski:2020pma} near a neutrino mass of 0.2 MeV. In 
the notation of Ref.~\cite{Falkowski:2020pma}, 
$\tilde\epsilon_L=c_{HNe}v^2/2\Lambda^2$ and the 
$\beta$ decay constraints extend to masses up to 
typical nuclear reaction 
energies of a few MeV.

Correspondingly, the strongest bound on the 
$c_{HN\mu}$ coupling transitions from the pion decay 
$\pi^+\to\mu^++\nu_R$~\cite{PhysRevD.36.2624} to the 
muon decay $\mu^-\to \nu_R+\bar\nu+ e^-
$~\cite{Fernandez-Martinez:2023phj} for neutrino 
masses around 2 MeV.

\begin{figure}
    \centering
\subfigure[]{
    \includegraphics[width=1\linewidth]{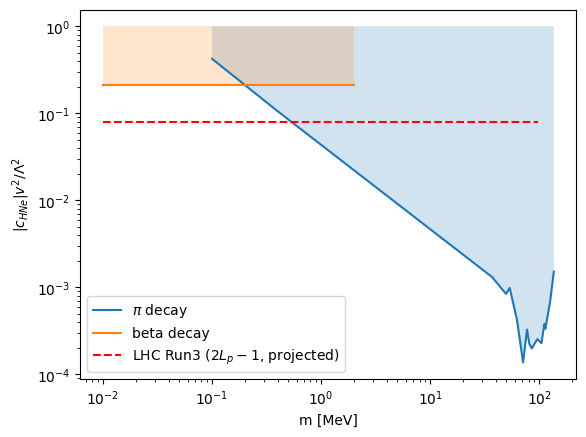}
}
\subfigure[]{
    \includegraphics[width=1\linewidth]{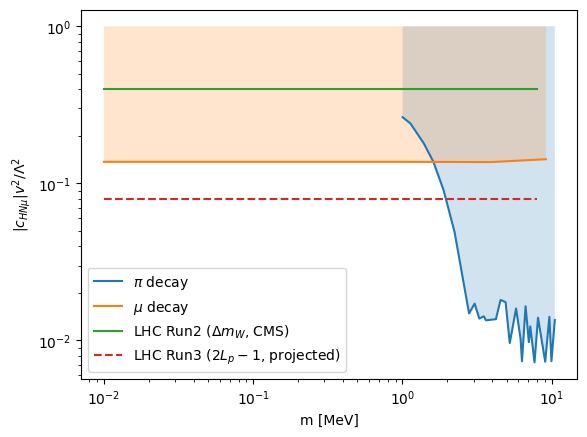}
}
\caption{
$95\%$ C.L. bounds on the probed Wilson coefficient from low-energy 
probes and from present and possible collider measurements. (a) 
Constraint on $|c_{HNe}|$ from nuclear $\beta$ decay (orange) and 
$\pi\!\to\! e\nu$ (blue). (b) Constraint on $|c_{HN\mu}|$ from 
muon (orange) and pion (blue) decays, and from the CMS measurement 
of$\Delta m_W\!\equiv\!m_{W^+}\!-\!m_{W^-}$ (green).  In both panels, the projected LHC Run 3 constraints from lepton angular measurements are shown as red dashed lines.
}
    \label{fig:other}
\end{figure}

These constraints are summarized in Fig.~\ref{fig:other}.  Current constraints from $W$-boson mass measurements are not competitive due to the $2\sigma$ difference between $m_{W^+}$ and $m_{W^-}$ observed by CMS, for which a $c_{HN\mu}$ coupling interpretation is excluded by muon decay measurements.

Future measurements should become more competitive, particularly if the mass difference is provided separately for each decay channel.  A measurement of the lepton decay angle in Run 3 of the LHC would improve on the constraints from muon decay by about a factor of two for right-handed neutrino masses below a couple of MeV.

\section{Conclusions}
\label{sec:conclusions}
We have investigated the collider phenomenology of a right-handed neutrino that has a mass below a GeV and interacts with the Standard Model through the dimension-six operator $\frac{c_{HNe}^i}{\Lambda^{2}}\,\bar\nu_{R}\gamma^{\mu}e_{R}\,\tilde H^{\dagger}iD_{\mu}H$. 
This interaction induces a parity-violating V+A component in $W\to\ell\nu$ decays, modifying both the kinematic distributions and the precision observables associated with the \(W\) boson. Working within the framework of the \(\nu\)SMEFT, we provided a quantitative analysis of the impact of such interactions on current and future collider measurements.

Revisiting the CDF Run II analysis showed that the presence of a \(W\to\ell\nu_R\) decay would shift the fitted \(W\)-boson mass upward by \(+76~\mathrm{MeV}\) for a Wilson coefficient near the perturbative bound, \(c_{HN\ell} v^{2}/\Lambda^{2}\simeq 1.2\). Such a coefficient would also induce differences between measurements of $m_{W^+}$ and $m_{W^-}$ at ATLAS and CMS. No such deviations are observed, and a 95\% CL bound of \(|c_{HN\ell}|v^{2}/\Lambda^{2}<0.27\) is inferred, thereby excluding this interpretation of the CDF measurement. 

We have proposed a probe of the \(\nu\)SMEFT using an angular observable based on the projection variable $2L_{p}-1\approx\cos\theta^*$, evaluated in the boosted regime where $p_T^W > 50~\mathrm{GeV}$. The reversed angular correlation induced by the V+A interaction relative to the Standard Model provides strong sensitivity to the model.  With the Run 3 dataset at $\sqrt{s}=13.6~\mathrm{TeV}$ and an integrated luminosity of $300~\mathrm{fb}^{-1}$, we find an expected limit of $c_{HNe}^i v^2/\Lambda^2<0.08$.

These bounds are competitive with those derived from low-energy probes such as nuclear $\beta$ decay, Michel parameters in muon decay, and pion decay. Importantly, for right-handed neutrino masses below a few MeV, future $W$-boson measurements can have dominant sensitivity. 

We have also assessed the potential impact on parton distribution functions, in particular through modifications to the $W$-charge asymmetry data used in PDF extractions. Our analysis indicates that the inclusion of the V+A component 
introduces only a modest shift in the relevant charge asymmetries relative to the shift in $m_W$. 

Looking forward, extending the angular analysis to the full HL-LHC dataset with $3~\mathrm{ab}^{-1}$ could further improve sensitivity. Additional gains could be achieved by moving to a higher $p_T^W$ regime and by constraining systematic uncertainties with a profile likelihood fit.  

In summary, the right-handed neutrino interpretation of the CDF \(W\)-mass measurement is ruled out by existing LHC measurements.  Measurements of the lepton decay angle could sharpen the constraints significantly in the near future. Together, our results outline a strategy for testing V+A interactions arising from light sterile neutrinos through $W$-boson measurements at current and future hadron colliders.

\begin{acknowledgments}
C.H. and C.P. acknowledge the Senior IPPP Fellowship, which funded this work.  R.A. and M.S. are supported by STFC under Grants No. ST/X003167/1 and No.
ST/X000745/1.
\end{acknowledgments}

%\bibliography{rhn}% Produces the bibliography via BibTeX.
\bibliography{rhn}

\end{document}